\documentclass[
aps,%
12pt,%
final,%
notitlepage,%
oneside,%
onecolumn,%
nobibnotes,%
nofootinbib,
superscriptaddress,%
noshowpacs,%
centertags]%
{revtex4}
\RequirePackage{graphicx}
\usepackage{epsfig}
\newcommand{\be}{\begin{eqnarray}}
\newcommand{\ee}{\end{eqnarray}}

\def\refitem#1{\relax}

\begin{document}
\title{Dynamical and  thermal descriptions in 
parton distribution functions.}

\author{\firstname{J.} \surname{Cleymans}}
\email{Jean.Cleymans@uct.ac.za}
\affiliation{UCT-CERN Research Centre and Department of Physics,
University of Cape Town 
}
\author{\firstname{G.I.} \surname{Lykasov}}
\email{lykasov@jinr.ru}
\author{\firstname{A.S.} \surname{Sorin}}
\email{sorin@theor.jinr.ru}
\author{\firstname{O.V.} \surname{Teryaev}}
\email{teryaev@theor.jinr.ru}
\affiliation{JINR Dubna,  
141980, Moscow region, Russia}

\begin{abstract}
We suggest a duality between the standard (dynamical) and statistical distributions of partons in 
the nucleons. The temperature parameter entering into the statistical
form for the quark distributions is estimated. 
It is found that this effective temperature is practically the same for the dependence on longitudinal and 
transverse momenta and, in turn, it is close to the freeze-out temperature in high energy 
heavy-ion collisions.
\end{abstract}

\maketitle

The description of hadron production using statistical models
has been pioneered several decades ago  by 
E.Fermi \cite{Fermi:1950}, I.Pomeranchuk \cite{Pomeran:1951}, L.D.Landau\cite{Landau:1953}
and  R.Hagedorn \cite{Hagedorn:1965}. 
The transverse momentum  spectrum of particles
produced in hadron-hadron collisions can be  been presented in the simple form
$\rho_h\sim\exp(-m_{ht}/T)$,
 where $-m_{ht}$ is the transverse mass of the hadron $h$ and $T$
is sometimes called the {\it thermal freeze-out} temperature.

As is well known, the statistical (thermal) models have  been applied successfully to describe hadronic 
yields produced in heavy-ion 
collisions (see, for example, \cite{Wheaton}-\cite{Sinyukov:02}
and references therein). 
The temperature obtained in these analysis is often referred to as {\it chemical freeze-out} temperature
and is consistently slightly larger than the thermal freeze-out temperature.
At the same time, the source of very fast thermalization is currently unknown and  
alternative or complementary possibilities to explain the thermal spectra are of much interest. 

Situations where statistical models have been  applied while the notion of statistical system 
was not obvious are not uncommon. In particular,  the statistical model was 
successfully applied to analyze deep inelastic lepton-nucleon scattering (DIS) 
\cite{Soffer1,Soffer2}, where the statistical form \cite{Cleymans:1988,Mac:1989,Zhang:2009} for distributions of
unpolarized and polarized quarks and gluons (partons) in a nucleon was used. 

This model may be compared with more standard parametrization of parton distributions based
on the Regge theory at low $x$ and quark counting rules at large $x$ \cite{Cap-Kaid}-\cite{NNPD}. 
It is especially interesting to find a counterpart and 
give a physical meaning of the  temperature 
introduced in the statistical model for the description of parton distribution functions.
This is the main goal of this paper. Making this comparison we suggest a concept of  
the {\it duality} between the statistical and
dynamical descriptions of the DIS and strong interactions of particles and explore its 
consequences and implications.     We do not consider the question of scaling violations for 
different values of $Q^2$, these will come into play when considering gluon or quark radiation and
need higher order diagrams to play a role. 

The  standard general form of parton distributions at the input energy scale $Q_0^2$,
 which is used in the perturbative QCD 
\cite{DGLAP} as the initial condition for the $Q^2$ evolution,
describing the experimental data on the DIS (see, e.g., \cite{Cap-Kaid})
is of the following general form:
\begin{equation}
xq(x)=A_s x^{-a_s}(1-x)^{b_s}~+~B_v x^{a_v}(1-x)^{b_v}~,
\label{def:F2in} 
\end{equation}
where the first term is a generic description of the sea quark distribution which is enhanced at low $x$,
according to the DIS experimental data, whereas the second term is due to the valence quark distribution. 
 Note that to fit the experimental data on the DIS a 
much more complicated form for $xq(x)$ is  used in the form of PDF's
\cite{CTEQ}-\cite{NNPD}.
        
In particular, the valence part at  $x<1/b_{q_v}$ can, to a good approximation, be written as: 
\be
xq_v(x)\simeq B_{q_v} x^{a_v}\exp\left(-x b_{q_v}\right)~.
\label{def:udstrexp}
\ee
Introducing the quantity ${\bar x}=1/b_{q_v}$ Eq.(\ref{def:udstrexp}) can be presented also
in the following form
\be
xq_v(x)\simeq B_{q_v} x^{a_v}\exp\left(-\frac{x}{{\bar x}}\right)~.
\label{def:udstrexpxm}
\ee
It would be interesting to find a link of the parameter ${\bar x}$ to the temperature $T$. 
Therefore, let us present a short review on the modern understanding of the statistical model of the 
hadron production and the nucleon structure functions. 

A statistical approach was applied in great detail to study the quark structure of nucleons in Ref.~\cite{Soffer1}, using 
earlier approaches as a starting point~\cite{Cleymans:1988,Mac:1989}.
The nucleon is viewed as a gas of massless partons (quarks, antiquarks, gluons) in equilibrium at
a given temperature in a finite size volume. 
According to the statistical approach, the quark distribution in  a proton $q(x)$ 
in the infinite momentum frame (IMF) at the input energy scale $Q_0^2$
is fitted as a function of the light cone 
variable $x$  in the following form:
\be
xq(x)=\frac{B_1 X_{0q}x^{b_1}}{\exp\left((x-X_{0q})/{\bar x}\right)+1}+
\frac{B_2 X_{0q}x^{b_2}}{\exp(x/{\bar x})+1}~,
\label{def:xqSoff}
\ee
where ${\bar x}$ plays the role of a {\it temperature inside the proton} and $X_{0q}$ is the chemical potential of the 
quark inside the proton. All the parameters are found from the description of the DIS.
The same form is suggested in \cite{Soffer2} for antiquarks $x{\bar q}(x)$ by changing the sign for the 
chemical potential $X_{0q}$.
Actually, Eq.(\ref{def:xqSoff}) is not the exact statistical form, this is similar to the Fermi-Dirac distribution.  
A meaning of  the factors $x^{b_1}$ and $x^{b_2}$ in 
Eq.(\ref{def:xqSoff}) is the same as the meaning of factors $x^{a_v}$ and $x^{-a_s}$ entered into Eq.(\ref{def:F2in}). 
They are dictated by the Regge behaviour at $x\rightarrow 0$ and the 
DIS experimental data which show some enhancement for the see quark distribution $xq_s(x)$ 
in comparison to the valence quark distribution $xq_v(x)$ at low $x$.

Note that the statistical weight of the quarks can be written in the form $exp((E_q-\mu)/T)$, where 
$E_q=(P \cdot p_q)/m$ is a quark energy in the nucleon rest frame, 
$p_q$ and $P$ are the four-momenta of quark and nucleon,  respectively, $m$ 
is the nucleon mass.\\
Passing to the infinite momentum frame the quark distribution in a nucleon over the longitudinal momentum
$p_z$ can be represented in the following statistical form \cite{Mac:1989}:
\be    
q(p_z)\sim \exp\left\{-\left(\frac{E\epsilon-Pp_z}{m T}-\frac{\mu}{T}\right)\right\}~,
\label{def:qpz}
\ee
where $E$, $P$ and $m$ are the  energy, the momentum and the mass of the  nucleon moving in the $z$-direction, $\epsilon$ 
and $p_z$ are 
the  energy and momentum in the $z$-direction of a parton in the nucleon, $T$ is the temperature and $\mu$ is the 
chemical potential. 
For massless partons  one  gets\\
\be    
q(x)\sim  \exp\left(-\frac{E_q}{T}\right)=\exp\left(-\frac{mx}{2T}\right)\equiv \exp(-\frac{x}{{\bar x}})~.
\label{def:apprqpx}
\ee
where $x=p_z/P$ is the longitudinal fraction of the parton momentum and $\bar{x} = 2T/m$. This form agrees with the 
result obtained in~\cite{Cleymans:1988,Mac:1989}. The inclusion of of the quark transverse momentum $k_{qt}$ results in
$E\epsilon-Pp_z\simeq m^2 x(1+k^2_{qt}/(x^2 m^2))$. It means that the relativistic invariant variable $x$ is replaced by
the variable $x(1+k^2_{qt}/(x^2 m^2))$ \cite{Efremov:2009ze}. 

This form is similar to Eq.(\ref{def:udstrexpxm}) if  ${\bar x}=1/b_v$. In this case the proton
temperature
is $T={\bar x}\cdot m/2=m/(2b_v(Q_0^2))$ for current quarks which are practically massless. 
Approximately one has for the valence 
$u$-quarks $b_{u_v}\simeq 2.5-3$ and for $d$-quarks $b_{d_v}\simeq 3.5-4$
at $Q_0^2=2-4$ (GeV$/c$)$^2$ \cite{Cap-Kaid,MRST}; therefore, $T\simeq 120-150$ MeV for the 
massless quarks.

The similarity of the thermal form for the quark distribution given by Eq.(\ref{def:apprqpx}) and  
the dynamic form for $q_v$ (Eq.(\ref{def:udstrexp}) maybe interpreted as  
{\it a duality of the thermal and dynamical} descriptions of the parton distribution in proton.
This property is by no means surprising as both parametrization are fitted to describe the same  
data. In Fig.1, the valence $u$-quark distribution in the proton as a function of $x$ is presented,
the solid curve corresponds to the CKMT parametrization \cite{Cap-Kaid}, see the second term of 
Eq.(\ref{def:F2in}) at $Q_0^2=4(GeV/c)^2$; the dotted line corresponds to the statistical (thermal) BS 
model \cite{Soffer2}, see the first term of Eq.(\ref{def:xqSoff}) at the same value of $Q_0^2$; 
the dash-dotted curve corresponds to the statistic parametrization given by Eq.(\ref{def:udstrexp}).    
Figure 1 shows that all three lines are very close to each other at $0.01\leq x\leq 0.4$.

 Let us compare the parameters of the statistical parametrization for $u_v(x)$ given by Eq.(\ref{def:udstrexp})
and the BS parametrization (Eq.(\ref{def:xqSoff})). 
Equation (\ref{def:xqSoff}) is not a thermal distribution. The parametrization
given by Eq.(\ref{def:udstrexp}) is  different . There is thus no contradiction in getting
50 MeV with one parametrization and 120 - 150 MeV with another one.
While the temperature has an evident dynamical counterpart 
$T=m/(2 b_v)$ , a similar relation for the chemical potential $X_{0q}$ of valence quarks is not obvious. 
The temperature $T=m/(2 b_v)$
entering into Eq.(\ref{def:udstrexp}) is about $120-150$ MeV, 
whereas the model of Ref.~\cite{Soffer2} suggests that $T$ is about $50$ MeV and the chemical potential for $u$-quarks
$X_{0u}\simeq 216$ MeV as first found in~\cite{Cleymans:1988,Mac:1989}.    

The quark distribution function  in a proton using the variable   $x$ and the transverse momentum $k_t$ can be represented
in the factorized form $f_q(x,k_t)=q(x)g(k_t)$ which of course cannot be valid at all values of $x$ 
\cite{Sehgal:1974bp}. This is supported by simulations within the lattice QCD \cite{LQCD} and by the 
observation of the so-called
``seagull'' effect \cite{Ammosov}, i.e., the weak $x$-dependence of the mean transverse momentum $<p_t>$ 
of hadrons produced in hadron inclusive reactions at low $x$, e.g., $x<0.5$ and the strong $x$-dependence 
of $<p_t>(x)$ at $x\rightarrow 1$, see for example \cite{LS:1991} and references therein.    

Developing the approach of the previous section one can fit the quark distribution $g(k_t)$ also 
in the statistical  form, i.e., 
\be
g(k_t)\sim \exp(-\frac{\epsilon_{kt}}{T})~,
\label{def:gkt}
\ee
where $\epsilon_{kt}=\sqrt{k_t^2+m_q^2}$ is the transverse energy of quarks in proton and 
$m_q$ is the quark mass.
For  massless quarks (the current quarks) $g(k_t)$ can be represented in the following form:
\be
g(k_t)\sim \exp(-\frac{k_t}{T})= ~\exp(-\frac{k_t}{<k_t>}).
\label{def:gktappr}
\ee
Applying here the same effective temperature  $T\simeq 120-150$ MeV, as for longitudinal momentum distribution,
we get similar results on $<k_t>$, which were obtained for valence current quarks 
(see e.g.\cite{Efremov:2009ze} and Ref. therein).
Actually, these values for $<k_t>$ of quarks in proton are used to get the 
inclusive transverse momentum spectra of hadrons produced in $p-p$ collisions at not large 
$p_t$, which were obtained in \cite{LS:1991}-\cite{LKSB:09} within the dual parton model (DPM) 
\cite{DPM} or the quark-gluon string model (QGSM) \cite{QGSM}.

Therefore, the transverse momentum distribution of partons in proton can also be described in the
statistical form with the same value of the temperature $T$ as the parton distributions
in proton over the longitudinal momentum or over its fraction $x$.  

Note that the longitudinal and transverse momentum dependences have been related in a model approach 
implying Lorentz (rotational) invariance \cite{Efremov:2009ze}

We analyzed the parton distribution functions in a proton using the statistical model \cite{Soffer1,Cleymans:1988,Mac:1989}. 
We suggest a duality principle which means a similarity of the thermal distributions of partons 
(quarks and gluons) and the dynamical
description of these partons. This duality allowed us to find an {\it effective temperature} 
$T \sim 120-150$ MeV for the massless quarks.
{\bf Concluding our study of this problem one can suggest another interpretation for this {\it effective temperature}
$T$ entered into the quark distributions given by Eq.(\ref{def:apprqpx}) and Eq.(\ref{def:gkt}) as the {\it effective widths}  
for the quark distributions over the longitudinal and transverse momentum, the values of which approximately are the same. 
}

The inclusive spectra of hadrons produced in $NN$ collisions at high energies are not described by the
thermal statistical distribution. However, using the thermal statistical form for quark distribution as a function 
of the longitudinal momentum fraction $x$ and the internal transverse momentum $k_t$ one can get a satisfactory 
description of the experimental data on these spectra.

It happens that the freeze-out temperature in central heavy-ion collisions at zero chemical 
potential {\bf or the {\it effective width} of the energy distribution of pions
}
has a similar value which was estimated in this paper for the 
massless quarks in a free nucleon   
\cite{Br-Munz:98}-\cite{LSST:09}.                                             
\vspace{0.25cm}
{\bf Acknowledgments}\\
The authors are grateful to A.Andronic, P.Braun-Munzinger, K.Bugaev, A.V.Efremov, L.L.Frankfurt,  
M.~Gazdzicki, S.B.~Gerasimov, M.Gorenstein, L.Jenkovscky, 
\frame{ A.B.~Kaidalov}, A.~Kotikov, C.Merino, A.~Parvan, V.B.Priezzhev, H.~Satz, A.~Sidorov, 
Yu.~Sinyukov, \frame{A.N.Sissakian} O.~Shevchenko, and J.~Soffer for very useful discussions. 
This work was supported in part by RFBR project No 11-02-01538-a.

%
\newpage
\begin{figure}[t]
\rotatebox{270}%
{\epsfig{file=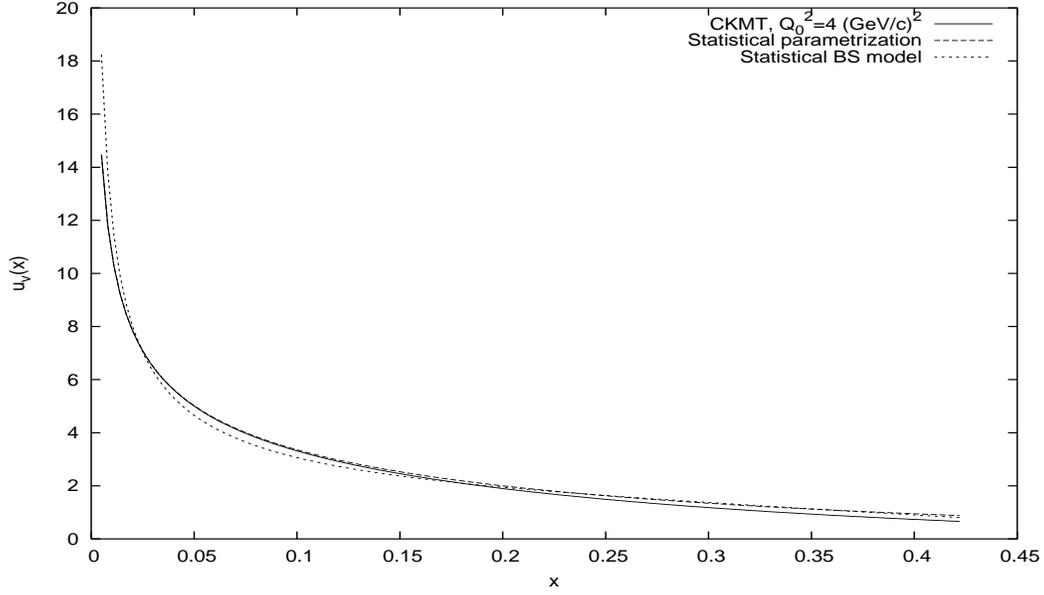,height=14.cm,width=8.cm }}
\caption[Fig.1] {The valence $u$-quark distribution in the proton as a function of the Bjorken variable $x$ at 
$Q_0^2=4(GeV/c)^2$. The solid curve is the CKMT parametrization \cite{Cap-Kaid}, the dashed
line corresponds to the statistical parametrization given by Eq.(\ref{def:xqSoff}) the 
dotted curve corresponds to the BS statistical (thermal) model suggested in \cite{Soffer1,Soffer2}.
}
\end{figure}
%
%
\newpage
\begin{center}
FIGURE CAPTIONS
\end{center}
\begin{enumerate}
\item
The valence $u$-quark distribution in the proton as a function of the Bjorken variable $x$ at 
$Q_0^2=4(GeV/c)^2$. The solid curve is the CKMT parametrization \cite{Cap-Kaid}, the dashed
line corresponds to the statistical parametrization given by Eq.(\ref{def:xqSoff}) the 
dotted curve corresponds to the BS statistical (thermal) model suggested in \cite{Soffer1,Soffer2}.
\end{enumerate}
\end{document}